\documentclass[12pt]{article}
\usepackage{graphicx}
\usepackage{graphics}
\usepackage{fancybox}
\usepackage{amsmath,float,latexsym,psfrag,epsf,epsfig,amssymb,
chicago,rotating}
\usepackage{color}
\usepackage{url}

\usepackage{fullpage}
\usepackage[ruled,linesnumbered]{algorithm2e}
\usepackage{algorithmicx}

\newcommand{\comment}[1]{}

\newcommand{\bfy}{\hbox{\boldmath$y$}}

\begin{document}

\begin{center}

{\Large \bfseries Bayesian inference for Gibbs random fields using composite likelihoods}
\vspace{5 mm}

{\large N. FRIEL\footnote{\texttt{nial.friel@ucd.ie}}} \\
{\textit{School of Mathematical Sciences and Complex Adaptive Systems Laboratory,\\ University College Dublin, Ireland.}}\\
\vspace{5 mm}

\today 

\vspace{5mm}

\end{center}

\bibliographystyle{mybib}

\begin{abstract}

\noindent Gibbs random fields play an important role in statistics, for example the autologistic model is commonly
used to model the spatial distribution of binary variables defined on a lattice. However they are complicated
to work with due to an intractability of the likelihood function. It is therefore natural to consider tractable 
approximations to the likelihood function. Composite likelihoods offer a principled approach to constructing 
such approximation. The contribution of this paper is to examine the performance of a collection of composite
likelihood approximations in the context of Bayesian inference. 


\end{abstract}

\section{Introduction}

Gibbs random fields play an important and varied role in statistics. The autologistic model is used to model the
spatial distribution of binary random variables defined on a lattice or grid \cite{bes74}. The exponential random 
graph model or $p^*$ model is arguably the most popular statistical model in social network analysis \shortcite{rob:pat:kal:lus07}.
Other application areas include biology, ecology and physics. 

Despite their popularity, Gibbs random fields present considerable difficulties from the point of 
view of parameter estimation, because the likelihood function is typically intractable for all but trivially small graphs. 
One of the earliest approaches to overcome this difficulty is the pseudolikelihood method \cite{bes72}, which approximates 
the joint likelihood function by the product of full-conditional distributions of all nodes. Indeed the pseudolikelihood
approximation is an example of a composite likelihood, that is, a likelihood approximation consisting of a product of a 
joint distribution of a lower dimensional variables, each of which can be normalised. It is natural to consider approximations
which refine pseudolikelihood by considering products of larger collections of variables. The purpose of this paper is to
consider such composite likelihood approximations. A similar study has been conducted by 
Okabayshi \textit{et al.} \citeyear{oka:john:gey11}, although from a likelihood inference perspective. As in this paper, they 
consider likelihood approximations consisting of a product of a joint distribution of collections of neighbouring variables. 
Using the recursion method of \cite{rev:pet04} we show that larger collections of variables can be used.

This paper is organised as follows. Section~\ref{sec:gibbs_ran_fields} outlines a description of Gibbs random fields, and in
particular the autologistic distribution. Composite likelihoods are introduced in Section~\ref{sec:comp_like}. Here we focus
especially on how to formulate conditional composite likelihoods for application to the autologistic model. We also focus on
the issue of calibrating the composite likelihood function for use in a Bayesian context. Section~\ref{sec:examples} illustrates
the performance of the various estimators for simulated data. The paper concludes with some remarks in 
Section~\ref{sec:conclusions}.

\section{Discrete-valued Markov random fields}
\label{sec:gibbs_ran_fields}

Discrete Markov random fields play an important role in several areas of statistics including spatial statistics and social
network analysis. The autologistic model, popularised by Besag \citeyear{bes72} which has the Ising model as a special case, is
widely used in analysis of binary spatial data defined on a lattice. The exponential random graph (or p$^*$) model is frequently
used to model relational network data. See \shortcite{rob:pat:kal:lus07} for an excellent introduction to this body of work. 

Let $y=\{y_1,\dots,y_n\}$ denote 
realised data defined on a set of nodes $\{1,\dots,n\}$ of a graph, where each observation $y_i$ takes values from some finite
state space. The likelihood of $y$ given a vector of parameters $\theta = (\theta_1,\dots,\theta_m)$ is defined as
\begin{equation}
 f(y|\theta) \propto \exp(\theta^T s(y)) := q(y|\theta),
\label{eqn:gibbs_like}
\end{equation}
where $s(y) = (s_1(y),\dots,s_m(y))$ is a vector of sufficient statistics. The constant of proportionality in 
(\ref{eqn:gibbs_like}), 
\[
 z(\theta) = \sum_{y\in Y} \exp(\theta^T s(y)),
\]
depends on the parameters $\theta$, and is a summation over all possible realisation of the Gibbs random field. Clearly, $z(\theta)$
is intractable for all but trivially small situations. This poses serious difficulties in terms of estimating the parameter vector
$\theta$. 

One of the earliest approaches to overcome the intractability of the (\ref{eqn:gibbs_like}) is the pseudolikelihood method \cite{bes75} 
which approximates the joint distribution of $y$ as the product of full-conditional distributions for each $y_i$,
\[
 f_{pseudo}(y) = \prod_{i=1}^n f(y_i|y_{-i},\theta),
\]
where $y_{-i}$ denotes $y\setminus\{y_i\}$. This approximation has been shown to lead to unreliable estimates of 
$\theta$, for example, \cite{ryd:tit98}, \shortcite{fri:pet:rev09}. This is in fact one of the earliest composite likelihood 
approximations, and we will outline work in this area further in Section~\ref{sec:comp_like}. Note also that Monte Carlo approaches 
have also been exploited to estimate the intractable likelihood, for example the Monte
Carlo maximum likelihood estimator of  Geyer and Thompson \citeyear{gey:tho92}. More recently, auxiliary variable approaches have been
presented to tackle this problem through the single auxiliary variable method \shortcite{mol:pet06} and the exchange algorithm  
\shortcite{Murray06}. 

The autologistic model, first proposed by Besag \citeyear{bes72}, is defined on a regular lattice of size $m\times m'$, where
$n=mm'$. It is used to model the spatial distribution of binary variables, taking values $-1,1$. The autologistic model is 
defined in terms of two sufficient statistics,
\[
 s_0(y) = \sum_{i=1}^n y_i, \;\;\; s_1(y) = \sum_{j=1}^n \sum_{i\sim j} y_i y_j,
\]
where the notation $i\sim j$ means that lattice point $i$ is a neighbour of lattice point $j$. 
Henceforth we assume that the lattice 
points have been indexed from top to bottom in each column and where columns are ordered
from left to right. For example, for a first order neighbourhood model where an interior point $y_i$ has neighbours 
$\{y_{i-m}, y_{i-1}, y_{i+1},y_{i+m}\}$. Along the edges of the lattice each point has either $2$ or $3$ neighbours. 
The full-conditional of $y_i$ can be written as
\begin{equation}
 p(y_i|y_{-i},\theta) \propto \exp\left( \theta_0 y_i + \theta_1 y_i( y_{i-m}+y_{i-1}+y_{i+1}+y_{i+m}) \right),
\label{eqn:full-con}
\end{equation}
where $y_{-i}$ denotes $y$ excluding $y_i$. As before, the conditional distribution is modified along the edges of the lattice.
The Hammersley-Clifford theorem \cite{bes74} shows the equivalence between the model defined in (\ref{eqn:full-con}) and in 
(\ref{eqn:gibbs_like}). 
Note that parameter $\theta_0$ controls the relative abundance of $-1$ and $+1$ values. The parameter $\theta_1$ controls the
level of spatial aggregation. When $\theta>0$, neighbouring values tend to take similar values, thereby yielding homogeneous
regions in the lattice. Note that the Ising model is a special case, resulting from $\theta_0=0$. 

\section{Composite likelihoods}
\label{sec:comp_like}

There has been considerable recent interest in composite likelihood methods in the statistics literature under such headings 
as pairwise likelihood methods \cite{not:ryd99}, composite likelihoods \cite{hea:lel98}, \cite{cox:rei04} and split-data 
likelihoods \cite{ryd94}. These concepts have long-standing antecedents such as Besag's pseudolikelihood \cite{bes75}. The 
basic idea is to work with a likelihood made up of factors, each of which corresponds to the joint probability function of a 
small number of variables, two in the case of pairwise likelihoods. 


Let us know return to the case where $y$ is a realisation from an autologistic distribution. Following the previous
section we denote $A = \{1,\dots,mm'\}$ as an index set for the lattice points. Following \shortcite{asun10} we
consider a general form of composite likelihood written as
\begin{equation}
 f(y|\theta) \approx \prod_{i=1}^C p(y_{A_i}|y_{B_i},\theta) := CL(y|\theta).
\label{eqn:comp_like_approx}
\end{equation}
Some special cases arise:
\begin{enumerate}
 \item $A_i=A$, $B_i=\emptyset$, $C=1$ corresponds to the full likelihood.
 \item $B_i=\emptyset$ is often termed \textit{marginal composite likelihood}.
 \item $B_i =A\setminus A_i$ is often termed \textit{conditional composite likelihood}.
\end{enumerate}
The focus of this paper is on conditional composite likelihoods. Note that the pseudolikelihood approximation is a
special of $3.$ where each $A_i$ is a singleton. We restrict each $A_i$ to be of the same dimension and in particular to correspond to contiguous
square 'blocks' of lattice points of size $k\times k$. In terms of the value of $C$ in case $3.$, an exhaustive 
set of blocks would result in $C=(m-k+1)\times(n-k+1)$. In particular, we allow the collection of blocks $\{A_i\}$
to overlap with one another. Finally, it is worth noting that in our context marginal distributions, $p(y_{A_i}|\theta)$, for index
sets $A_i$, are rarely, if ever, available. Hence we don't consider marginal composite likelihoods in this context.

Our interest here is to compare the performance of estimation of $\theta$ using the conditional composite likelihood
and especially to understand the statistical efficiency as the block size $k$ increases. 
It should also be evident that the computational complexity of this approximation will increase dramatically with $k$,
the size of the blocks. Consequently our interest here is to explore the trade-off between using a larger block size,
with a smaller number of blocks $C$ in (\ref{eqn:comp_like_approx}).

\subsection{Computing full-conditional distributions of $A_i$}

The conditional composite likelihood which we described above relies on evaluating
\begin{equation}
 p(y_{A_i}|y_{-A_i},\theta) = \frac{\exp\left( \theta_0 s_0(y_{A_i}) + s_1(y_{A_i}|y_{-A_i}) \right)}{z(\theta,y_{A_i})},
\label{eqn:gen_recursions}
\end{equation}
where
\[
 s_0(y_{A_i}) = \sum_{j\in{A_i}} y_j, \;\;\; s_1(y_{A_i}|y_{-A_i}) = \sum_{j\in{A_i}}\sum_{l\sim j} y_l y_j.
\]
Also the normalising constant now includes the argument $y_{A_i}$ emphasising that it involves a summation over all possible 
realisations of sub-lattices defined on the set $A_i$ and conditioned on the realised $y_{-A_i}$. First we describe an approach 
to compute the overall normalising constant for a lattice, without any conditioning on a boundary. 

Generalised recursions for computing the normalizing constant of general factorisable models such as the autologistic models have
been proposed by \citeN{rev:pet04}, generalising a result known for hidden Markov Models 
(e.g. \citeNP{zucchini-guttorp1991,scott2002}). This method applies to autologistic lattices with a small number of rows, up to 
about 20, and is based on an algebraic simplification due to the reduction in dependence arising from the Markov property. It 
applies to un-normalized likelihoods that can be expressed as a product of factors, each of which is dependent on only a subset 
of the lattice sites. We can write $q(y|\theta)$ in factorisable form as
\begin{displaymath}
  q(y|\theta) = \prod_{i=1}^{n} q_i(\bfy_i|\beta),
\end{displaymath}
where each factor $q_i$ depends on a subset $\bfy_i = y_i,y_{i+1},\dots,y_{i+m}$ of $y$, where $m$ is 
defined to be the \emph{lag} of the model. We may define each factor as
\begin{equation}
  q_i(\bfy_i,\beta) = \exp\{ \theta_0 y_i + \theta_1 y_i(y_{i+1}+y_{i+m})\}
  \label{eqn:qi}
\end{equation}
for all $i$, except when $i$ corresponds to a lattice point on the last row or last column, in which case $y_{i+1}$ or $y_{i+m}$, 
respectively, drops out of~(\ref{eqn:qi}).

As a result of this factorisation, the summation for the normalizing constant,
\begin{displaymath}
z(\theta) = \sum_{y}\prod_{i=1}^{n} q_i(\bfy_i|\theta)
\end{displaymath}
can be represented as
\begin{equation}
z(\theta) = \sum_{y_{n}} q_{n}(\bfy_{n}|\theta) \sum_{y_{n-1}} q_{n-1}(\bfy_{n-1}|\theta) \dots  \sum_{y_1}
q_1(\bfy_1|\theta)
\label{eqn:z_theta}
\end{equation}
which can be computed much more efficiently than the straightforward summation over the $2^n$ possible lattice realisations. Full 
details of a recursive algorithm to compute the above can be found in \citeN{rev:pet04}. Note that this algorithm was extended 
in \cite{fri:rue07} to also allow exact draws from $p(y|\theta)$

The minimum lag representation for an autologistic lattice with a first order neighbourhood occurs for $r$ given by the smaller 
of the number of rows or columns in the lattice. Identifying the number of rows with the smaller dimension of the lattice, the 
computation time increases by a factor of two for each additional row, but linearly for additional columns.
It is straightforward to extend this algorithm to allow one to compute the normalising constant in (\ref{eqn:gen_recursions}),
so that the summation is over the variables $y_{A_i}$ and each factor involves conditioning on the set $y_{-A_i}$.

\subsection{Bayesian composite likelihoods}
\label{sec:bayes:cl}

The focus of interest in Bayesian inference is the posterior distribution
\[
 p(\theta|y) \propto f(y|\theta) p(\theta).
\]
Our proposal here is to replace the true likelihood $f(y|\theta)$ with a conditional composite likelihood approximation,
leading us to focus on the approximated posterior distribution
\[
 p^*(\theta|y) \propto CL(y|\theta) p(\theta). 
\]
Surprisingly, there is very little literature on the use of composite likelihoods in the Bayesian setting, although 
Pauli \textit{et al.} \citeyear{pauli11} present a discussion on the use of conditional composite likelihoods. Indeed this
paper suggests, following \cite{lin88}, that a composite likelihood should take the general form
\begin{equation}
 f(y|\theta) \approx \prod_{i=1}^C p(y_{A_i}|y_{B_i},\theta)^{w_i},
\label{eqn:comp_like_approx1}
\end{equation}
where $w_i$ are positive weights. In all experiments carried out here, we assume that $w_i=1$, and empirically we
observe that non-calibrated composite likelihood leads to an approximated posterior distribution with substantially
lower variability than the true posterior distribution, leading to overly precise precision about posterior parameters.

\section{Examples}
\label{sec:examples}

Here we simulated $20$ realisations from a first-order Ising model all 
defined on a $16\times 16$ lattice, with a single interaction parameter $\theta=0.4$, which is close to the critical phase 
transition beyond which all realised lattices takes either value $+1$ or $-1$. This parameter setting is the most challenging
for the Ising model, since realised lattices exhibit strong spatial correlation around this parameter value. 
For a lattice of this dimension it is possible to exactly calculate 
the normalising constant $z(\theta)$. Using a fine grid of $\{\theta_i\}$ values, the right hand side of:
\[
 p(\theta_i|y) \propto \frac{q(y|\theta_i)}{z(\theta_i)} p(\theta_i),\;\; i=1,\dots,n. 
\]
can be evaluated exactly. Summing up the right hand side yields an estimate of the evidence, $p(y)$, which is the constant of
normalisation for the expression above and which in turn can be used to give
a very precise estimate of $p(\theta|y)$. This serves as a ground truth against which 
to compare with the posterior estimates of $\theta$ using the various composite likelihood estimators. Exact calculation of $z(\theta)$ and the approach
described above to get a precise estimate of the evidence relies on algorithms developed in \cite{fri:rue07}. For this experiment,
we choose a uniform $[-10,10]$ prior for $\theta$ . 

\begin{figure}
 \begin{center}
 \includegraphics[width=8cm]{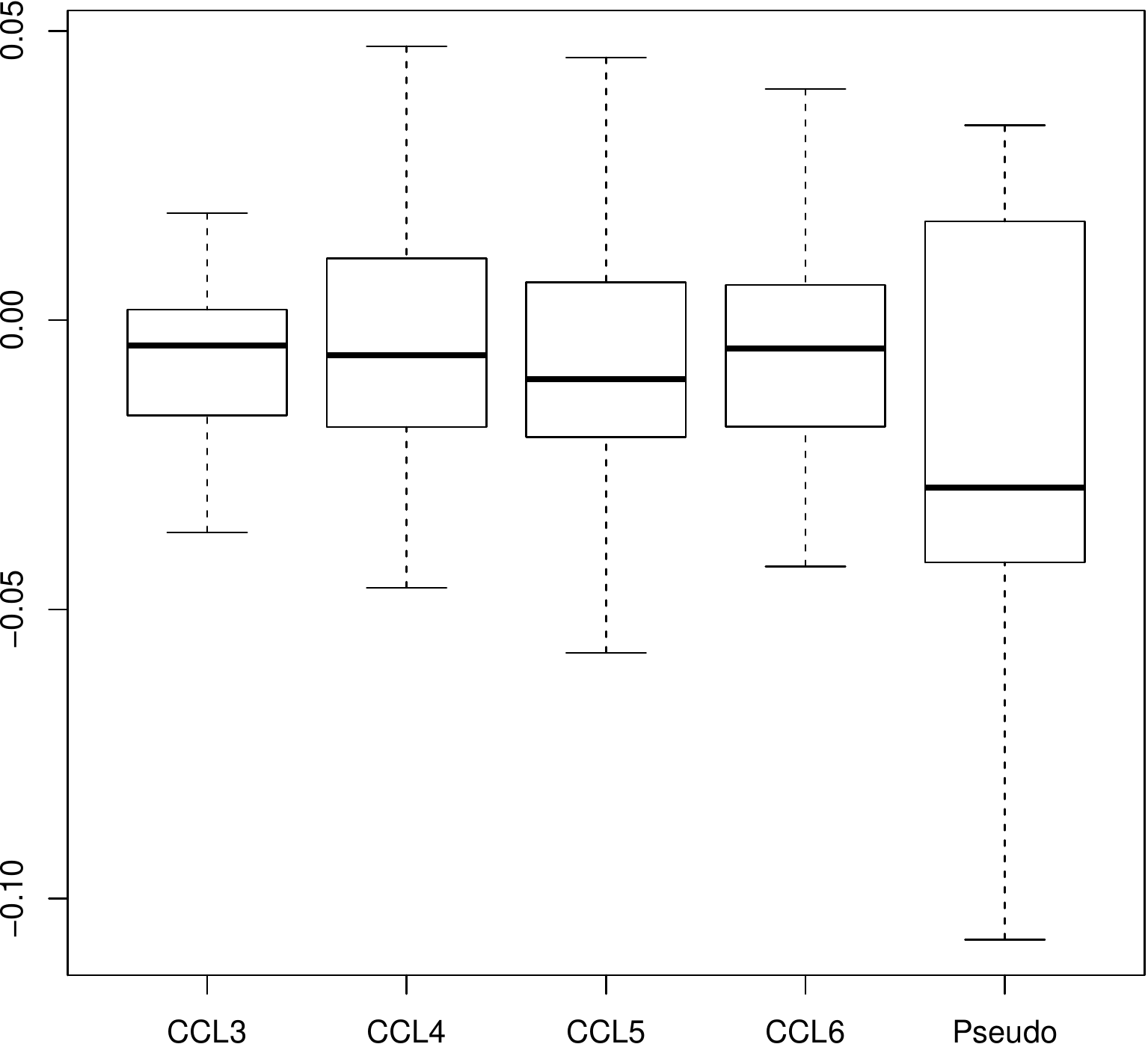}
\end{center}
\caption{$16\times 16$ lattices: Boxplot displaying the bias of the estimate of $\theta$ for $20$ datasets corresponding to each of $4$ 
conditional composite likelihood approximations, CCL$_3$, CCL$_4$, CCL$_5$ and CCL$_6$ (corresponding to block size 
$3\times 3$, $4\times 4$, $5\times 5$, $6\times 6$, respectively) and also pseudolikelihood estimator. 
Note that the computational time for each of the composite likelihood approximations was held constant. }
\label{fig:boxplot_16}
\end{figure}

In terms of MCMC implementation, $5000$ iterations were used with a burn in period of $1,000$ iterations for each dataset. 
Computation was carried out on a desktop PC with a $3.33$Ghz processor and with $4$Gb of memory. 
Computation time for each of the different composite likelihoods was approximately constant and took
$0.004$ second of CPU time per iteration. This was achieved by exhaustively using
all $3\times 3$ blocks in the CCL$_3$ approximation, $40\%$ of all $4\times 4$ blocks, $20\%$ of all $5\times 5$ blocks and
$10\%$ of all $6\times 6$ blocks in the CCL$_4$, CCL$_5$ and CCL$_6$ approximations, respectively. The results for simulations
involving the $16\times 16$ lattices are displayed in Figure~\ref{fig:boxplot_16}. Here each of the conditional composite 
likelihood methods perform better than pseudolikelihood. Each of the composite likelihood approximations performed equally 
well, although the CC$_4$, CCL$_5$ and CCL$_6$ display larger spread than CCL$_3$. 

It is apparent from Table~\ref{tab:post_var16} that the estimated posterior variance of $\theta$ for each of the approximations are
generally lower than the true posterior variances. In fact the conditional composite likelihood approximations lead to posterior
variance estimates than are smaller by a factor of $10$. This strongly suggests that the conditional composite likelihoods need
to be calibrated in some form. 
\begin{table}[htp]
\centering
\begin{tabular}{cccccc}
\hline
$CCL_3$ & $CCL_4$ & $CCL_5$ & $CCL_6$ & Pseudo & True \\
$2.1\times 10^{-4}$ & $3.0\times 10^{-4}$ & $3.3\times 10^{-4}$ &
$4.3\times 10^{-4}$ & $2.1\times 10^{-3}$ & $1.6\times 10^{-3}$ \\
\hline
\end{tabular}
\caption{$16\times 16$: Average posterior variance for $\theta$ for each of $20$ datasets.} 
\label{tab:post_var16}
\end{table}

In a similar vein to the previous experiment, here we examined the performance of the various approximations on simulated
data defined on larger $50\times 50$ lattice. In this instance we can't analytically compute the true likelihood, however 
here we used the 
exchange algorithm \cite{Murray06} to generate draws from the target posterior, from a very long MCMC run. The simulation study 
was otherwise similar in every other respect. Computation time was approximately $0.1$ second per iteration of the MCMC algorithm
using the various composite likelihood approximations. 
Here the performance of the conditional composite likelihood was again similar to each other, and better generally, in terms of 
lower bias, than posterior mean estimation using the pseudolikelihood approximation. Here, similar to the previous experiment,
we see in Table~\ref{tab:post_var50} that the posterior variance based on the various the conditional composite likelihood are
considerably small, than that estimated by the exchange algorithm. 

\begin{figure}
 \begin{center}
 \includegraphics[width=8cm]{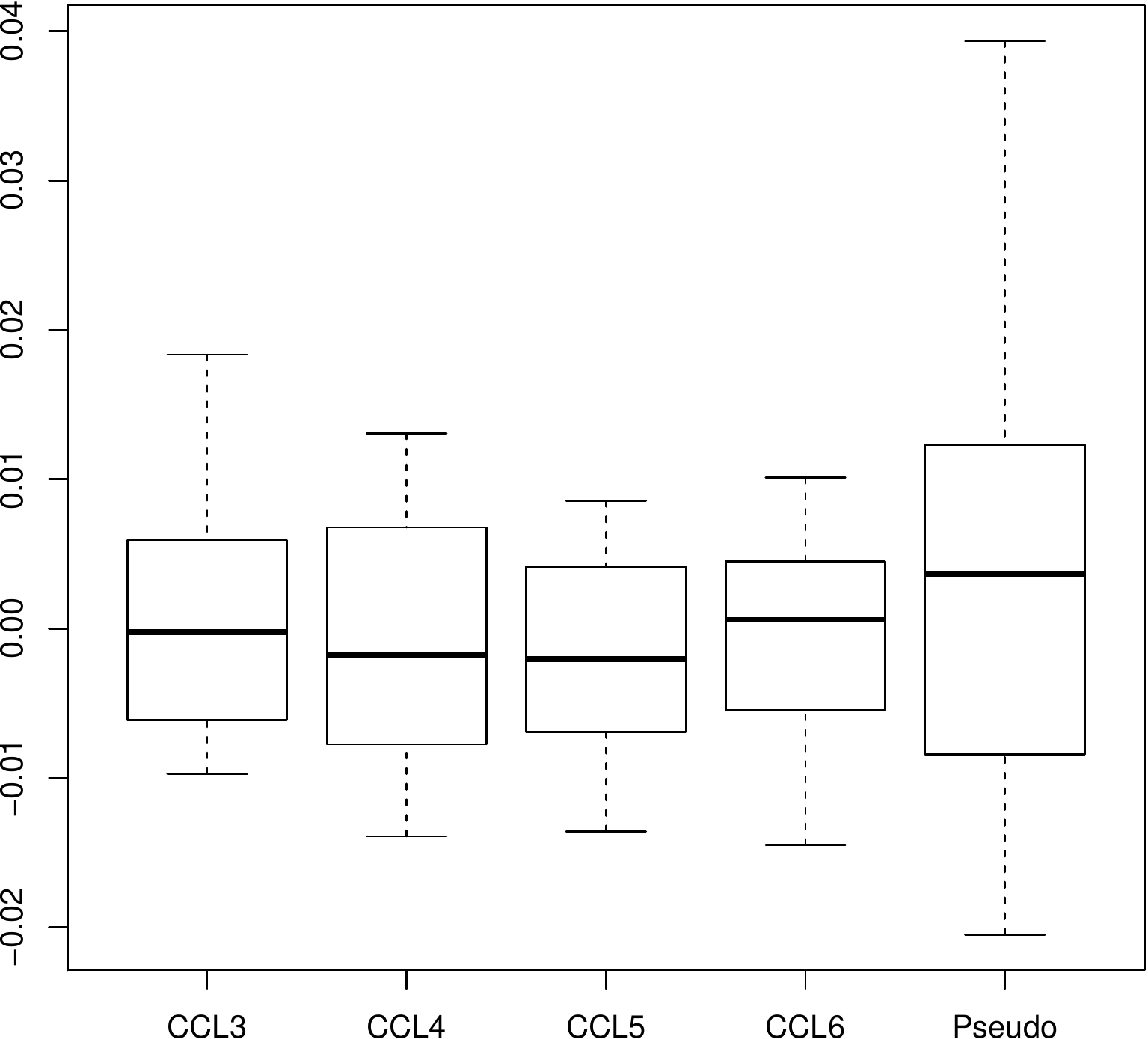}
\end{center}
\caption{$50\times 50$ lattices: Boxplot displaying the bias of the estimate of $\theta$ for $20$ datasets corresponding to each of $4$ 
conditional composite likelihood approximations, CCL$_3$, CCL$_4$, CCL$_5$ and CCL$_6$ (corresponding to block size 
$3\times 3$, $4\times 4$, $5\times 5$, $6\times 6$, respectively) and also pseudolikelihood estimator. 
Note that the computational time for each of the composite likelihood estimator was held constant. }
\label{fig:boxplot_50}
\end{figure}

\begin{table}[htp]
\centering
\begin{tabular}{cccccc}
\hline
$CCL_3$ & $CCL_4$ & $CCL_5$ & $CCL_6$ & Pseudo & True \\
$1.5\times 10^{-3}$ & $3.3\times 10^{-5}$ & $2.2\times 10^{-5}$ &
$2.6\times 10^{-5}$ & $2.1\times 10^{-4}$ & $1.3\times 10^{-2}$ \\
\hline
\end{tabular}
\caption{$50\times 50$: Average posterior variance for $\theta$ for each of $20$ datasets.} 
\label{tab:post_var50}
\end{table}

\subsection{Why is the posterior variance of estimators based on composite likelihoods overly precise?}

The results of this section suggest that using conditional composite likelihoods leads to considerably underestimated
posterior variances. A possible explanation for this behaviour may be due to a type of 'annealing' effect where the
true likelihood is replaced by a powered version of it, leading to an overly concentrated likelihood function. Here the
true likelihood $f(y|\theta)$ is replaced by $\prod_{i=1}^C p(y_{A_i}|y_{A\setminus A_i},\theta)^{w_i}$. Suppose that
$w_i=1$ for all $i$ (as is the case in all of the experiments carried out here) and suppose further that $C$ is large, 
whereby potentially many blocks overlap. In this scenario the set of interactions in the true likelihood will be a
subset of all the interactions in the conditional composite likelihood, due to the overlapping blocks, and this will
in turn lead to an annealing of the true likelihood function. 

\section{Conclusion}
\label{sec:conclusions}

This paper has illustrated the important role that composite likelihood approximations can play in the statistical
analysis of Gibbs random fields, and in particular in the Ising and autologistic models in spatial statistics. This
paper has focused on the use of conditional composite likelihoods, based on tractable full-conditional distributions 
over blocks of lattices points and shows much promise. However it is evident that the posterior distribution of the
interaction parameter $\theta$ is too concentrated and therefore underestimates the posterior variance. An important
research question is to ask how to correctly calibrate the conditional composite likelihood so that it achieves
the correct variance. 

The computational complexity of this approximation increases dramatically as the size of blocks increases and our study 
here shows that efficient parameter estimation can result by considering conditional composite likelihoods based on a 
subset of possible blocks, thereby reducing computation time. For future work it would be interesting to understand how
composite likelihoods can be usefully employed for more challenging Markov random fields models.  

\paragraph*{Acknowledgements:} Nial Friel's research was supported by a Science Foundation Ireland 
Research Frontiers Program grant, 09/RFP/MTH2199.

\bibliography{hmrf}

\end{document}